\documentclass[pre,aps,twocolumn,showpacs]{revtex4}
\usepackage{amssymb}
\usepackage{amsmath,bm}
\usepackage{graphicx}
\usepackage{color}

\begin{document}

\title{Passive scalar transport in peripheral regions of random flows}

\author{A. Chernykh$^{1,2}$ and V. Lebedev$^{3,4}$ \\
$^1$ Institute of Automation and Electrometry SB RAS \\
 $^2$ Novosibirsk State University \\
 $^3$ Landau Institute for Theoretical Physics RAS \\
 $^4$ Moscow Institute of Physics and Technology}

\begin{abstract}

We investigate statistical properties of the passive scalar mixing in random (turbulent)
flows assuming weakness of its diffusion. Then at advanced stages of the passive scalar
decay its unmixed residue is concentrated primarily in a narrow diffusive layer near the
wall and its transport to bulk goes through the peripheral region (laminar sublayer of
the flow). We conducted Lagrangian numerical simulations of the process for different
space dimensions $d$ and revealed structures responsible for the transport that are
passive scalar tongues pulled from the diffusive boundary layer to bulk. We investigated
statistical properties of the passive scalar and of the passive scalar integrated along
the wall. Moments of both objects demonstrate scaling behavior outside the diffusive
boundary layer. We propose an analytical scheme for the passive scalar statistics,
explaining the features observed numerically.

\end{abstract}

\pacs{05.10.-a, 05.20.-y, 05.40.-a, 47.27.-i}

\maketitle

\section{Introduction}
\label{sec:intro}

Stochastic dynamics of such scalar fields, as temperature or concentration of pollutants,
in random (turbulent) flows is of great importance in different physical contexts, from
cosmology to micro-fluidics. If the back reaction of the field to the flow is negligible
then the field is called passive scalar. We consider the passive scalar in random flows,
where the flow velocity varies randomly in time. Theoretical examination of dynamical and
statistical properties of the passive scalar in random flows goes back to classical works
of Obukhov and Corrsin \cite{obukhov,corrsin}, where a phenomenology for the passive
scalar statistics in turbulent flows was developed in spirit of the Kolmogorov scheme
\cite{kolmogorov}. Modern understanding of the passive scalar statistics in turbulent
flows is reflected in Refs. \cite{96Sre,00War,05SS,08MDW}, see also the books
\cite{MY,lesieur,Frisch}. The mixing problem for the passive scalar is investigated for
chaotic flows as well, see the book of Ottino \cite{Ottino}. An interesting example of a
random flow is the so-called elastic turbulence, discovered by Groisman and Steinberg
\cite{00GS} in polymer solutions. Observations of the passive scalar statistics in the
elastic turbulence were reported in Ref. \cite{05KS}.

In 90th, a series of theoretical works devoted to the passive scalar statistics was done
in the framework of the so-called Kraichnan model where the turbulent flow is assumed to
be short correlated in time and possessing a definite scaling. The works enabled one to
reveal general features of the passive scalar statistics in turbulent flows including the
so-called anomalous scaling and intermittency \cite{95GK,95SS,95CFKL}, see also the
surveys \cite{00SS,01FGV}. However, the approach implies spacial homogeneity of the flow
statistics and is not, consequently, directly applicable to regions near walls.

In this paper we investigate the passive scalar statistics in peripheral regions of a
vessel where the developed (high-Reynolds) turbulence is excited. Speaking about the
peripheral regions of the turbulent flows, we imply a laminar (viscous) sublayer formed
near walls where the velocity field can be regarded as smooth, it varies on distances of
the order of the thickness of the sublayer. However, the velocity remains a random
function of time there. Some laminar boundary layer is characteristic also of the elastic
turbulence \cite{07BSS}. The passive scalar statistics in the peripheral region is
determined by a complicated interplay of its diffusion and random advection in highly
anisotropic situation caused by presence of the walls.

We are interested in advanced stages of the passive scalar decay assuming that the Peclet
or the Schmidt number is large. In the case the unmixed fraction of the passive scalar is
located mainly in a narrow diffusive layer near the wall, thinner than the thickness of
the peripheral region \cite{03CLa}. Then the passive scalar transport to bulk goes
through the peripheral region outside the diffusive layer. Just this peripheral region
plays a crucial role in formation of statistical characteristics of the passive scalar
transport. The same reasoning can be applied to a stationary case related, say, to a
permanent heat flow going from the walls through the periphery region to bulk. Moreover,
fast chemical reactions can be analyzed inside the same scheme, see Ref. \cite{03CLb}. A
theoretical approach to the problem was developed in Ref. \cite{04LT}, principal
predictions of the theory were confirmed by experiment, see Ref. \cite{04BSS}.

To check the theoretical predictions in detail, we conducted extensive numerical
simulations of the problem based on Lagrangian dynamics of particles representing the
passive scalar. Aiming to establish main qualitative properties of the passive scalar
transport in the peripheral regions, we focused on the $2d$ (two-dimensional) case.
However, a big advantage of the Lagrangian scheme is an ability to extend the approach
without essential troubles to higher dimensions. We performed simulations for the space
dimensionalities $d=3\div5$ to establish universality of the passive scalar statistical
behavior in $2d$ and to reveal features characteristic of higher dimensions. We used a
scheme with permanent injection of particles near the wall that produces statistically
homogeneous in time statistics of the passive scalar. However, our conclusions are
correct for the decaying case as well because of adiabaticity: events responsible for the
passive scalar transport to bulk are much faster than the average passive scalar decay.

The obtained numerical data can be used to compute averages characterizing the passive
scalar statistics. First of all, we found moments of the passive scalar at different
separations from the wall. The data show an existence of a well pronounced diffusive
layer where the passive scalar is mainly concentrated, in accordance with the theoretical
expectations formulated in Ref. \cite{04LT}. Outside the diffusive layer the passive
scalar moments demonstrate scaling behavior with exponents deviating from ones proposed
in Ref. \cite{04LT} where diffusion was assumed to be negligible outside the diffusive
layer. We checked that the deviations are related to diffusion, indeed. The situation
resembles the passive scalar statistics in the Batchelor velocity field on scales larger
than the pumping length where diffusion appears to be relevant \cite{07CKL}, that
corrects the diffusionless behavior examined in Ref. \cite{99Lys}. Next, we introduced
the passive scalar integrated along a surface parallel to the wall. The diffusion along
the wall drops from the equation for the object. We demonstrated numerically that outside
the diffusive layer moments of the integral passive scalar have well pronounced scaling
behavior, as a function of separation from the wall. We found the corresponding scaling
exponents for moments with degrees $n=1\div6$ in space dimensionalities $d=2\div5$. The
moments exhibit an anomalous scaling signalling about strong intermittency of the passive
scalar statistics.

The simulations enabled us to reveal objects lying behind the intermittency. The objects
are tongues of the passive scalar pulled from the diffusive layer towards bulk. The
tongue cross section diminishes as the separation from the wall increases (due to
increasing the velocity component perpendicular to the wall). That explains why diffusion
can play an essential role even in the region outside the diffusive layer. The subsequent
tongue evolution including tongue folds produces long-living structures possessing
complex shape. Sometimes the tongues are pulled so strongly that push irreversibly a
passive scalar portion away from the wall. Just this mechanism is responsible for the
passive scalar transport to bulk, that naturally explains its strong intermittency.

We suggest a theoretical scheme for explanation of the passive scalar statistics, based
on smallness of the passive scalar correlation length along the wall outside the
diffusive layer. The scheme enables one to find explicit expressions for scaling
exponents characterizing different objects. A comparison of the theoretical predictions
with numerics shows that they agree satisfactory. Some preliminary results of our work
were published in Ref. \cite{08CL}.

The structure of the paper is as follows. In the second section we present our
theoretical approach concerning the passive scalar dynamics and statistics in the
peripheral region and propose a scheme giving the scaling exponents. In the third section
we explain our computational scheme, present computed moments of the passive scalar and
integral passive scalar, give a description of the passive scalar tongues, and compare
our numerical results with theory. In Conclusion we give an outline of our results, their
possible applications and directions of future investigations.

\section{Theoretical Desciption}
\label{sec:basic}

We consider the passive scalar statistics in peripheral regions of a random flow, that
are regions near boundaries (walls). Our principal example is the viscous (laminar)
boundary layer of the developed high-Reynolds hydrodynamic turbulence (see, e.g., the
book \cite{MY}). However, our approach can be applied to other situations. For example,
one can think about the peripheral region of the elastic turbulence \cite{00GS}. The only
feature relevant for us is smoothness of the flow in the boundary layer, whereas the
velocity varies randomly in time there.

We designate the passive scalar field as $\theta$. It can represent both, temperature
variations or concentration of pollutants. The passive scalar evolution (decay) in an
external flow is described by the advection-diffusion equation
 \begin{equation}
 \partial_t\theta+\bm v\nabla\theta
 =\kappa \nabla^2 \theta ,
 \label{basicpass}
 \end{equation}
where $\bm v$ is the flow velocity and $\kappa$ is the diffusion (thermodiffusion)
coefficient. Below, the fluid is assumed to be incompressible (that is the flow is
divergentless, $\nabla\bm v=0$). The equation (\ref{basicpass}) implies that there are no
sources of the passive scalar in the volume. However, we do not exclude a passive scalar
flux from the vessel walls.

The equation (\ref{basicpass}) has to be supplemented by boundary conditions at the wall.
If $\theta$ is the density of pollutants and the wall is impenetrable for the pollutants
then the gradient of $\theta$ in the direction perpendicular to the wall is zero near the
wall, that corresponds to zero pollutant flux to the boundary. In the case we deal with
the passive scalar decay, leading ultimately to its homogeneous distribution in space. If
$\theta$ is temperature then its gradient in the direction perpendicular to the wall can
be non-zero, that corresponds to finite heat flux through the boundary (from the wall).
If walls are made of a well heat conducting material then a value of $\theta$
(temperature) has to be regarded as fixed at the boundary.

We assume that the Peclet or the Schmidt number is large (that is the diffusion
coefficient $\kappa$ is small in comparison with the fluid kinematic viscosity $\nu$).
Then, as it was demonstrated in the work \cite{04LT}, the passive scalar dynamics in the
peripheral region is slow in comparison with the velocity dynamics. Therefore the passive
scalar is rapidly mixed in bulk (for a time that can be estimated as an inverse Lyapunov
exponent of the flow) that leads to a homogeneous spacial distribution of the passive
scalar, $\theta= \mathrm{const}$. The subsequent passive scalar evolution is related
mainly to the peripheral regions that supply bulk by passive scalar fluctuations. We
assume that bulk can be treated as a big reservoir, then the bulk homogeneous value of
$\theta$, $\theta_b$, can be regarded as independent of time. Below, we imply that the
passive scalar field is shifted by $\theta_b$, that leads to the condition $\theta\to0$
as we pass from the periphery to bulk.

\subsection{Correlation Functions}

Statistical properties of the passive scalar can be described in terms of its correlation
functions
 \begin{equation}
 F_n(t,\bm r_1,\dots,\bm r_n)=
 \langle\theta(t,\bm r_1)\dots
 \theta(t,\bm r_n)\rangle ,
 \label{pe9}
 \end{equation}
where angular brackets mean averaging over large times (larger than the velocity
correlation time). Since the velocity tends to zero at approaching the wall and the
molecular diffusion is assumed to be weak, the passive scalar dynamics, determined by an
interplay of the advection and diffusion, is slower than the velocity dynamics in the
peripheral region. Therefore at investigating the passive scalar dynamics the velocity
can be regarded as short correlated in time and closed equations can be derived for the
passive scalar correlation functions (see, e.g., \cite{01FGV,04LT})
\begin{eqnarray}
 \partial_t F_n
 =\kappa\sum_{m=1}^n\nabla_m^2F_n \qquad \qquad
 \nonumber \\
 +\sum_{m,k=1}^n  \sum_{\alpha\beta}
 \partial_{m\alpha} \left[D_{\alpha\beta}(\bm r_m,\bm r_k)
 \partial_{k\beta}F_n\right],
 \label{pe10}
 \end{eqnarray}
where the object $D_{\alpha\beta}$ is expressed via the pair velocity correlation
function as
 \begin{equation}
 D_{\alpha\beta}(\bm r_1,\bm r_2)=
 \int_0^\infty d t'\,
 \langle v_\alpha(t+t',\bm r_1)
 v_\beta(0,\bm r_2)\rangle \,.
 \label{pe5}
 \end{equation}
Here, again, angular brackets mean averaging over times larger than the velocity
correlation time.

The structure of the equation (\ref{pe10}) is quite transparent: the evolution of the
passive scalar correlation functions is determined by the molecular diffusion (the first
term in the right-hand side) and by the eddy diffusion (the second term in the right-hand
side). Therefore the quantity $D_{\alpha\beta}$ can be called the eddy diffusion tensor,
since it describes diffusion of the passive scalar related to the random flow. This
effect can be compared with the turbulent diffusion of the passive scalar in turbulent
flows in bulk on scales larger than the viscous length. However, in our case the eddy
diffusion tensor $D_{\alpha\beta}$ is associated with a smooth flow, and can be used for
description of the passive scalar dynamics on scales smaller than the turbulent viscous
length.

We assume that the walls of the vessel are smooth and that the boundary layer width is
much less than the curvature radii of the wall. Then it can be treated as flat in the
main approximation. Let us introduce a reference system with the $Z$-axis perpendicular
to the wall and assume that the fluid occupies the region $z>0$. Smoothness of the
velocity leads to the proportionality laws $v_x,v_y\propto z$ and $v_z\propto z^2$ for
the velocity components along and perpendicular to the wall, respectively. The laws are
consequences of the velocity smoothness, of the non-slipping boundary condition $\bm v=0$
at the wall, and of the incompressibility condition $\nabla\bm v=0$.

Below we assume that the velocity statistics is homogenous in time, and also assume its
homogeneity along the wall. Due to the assumed homogeneity, velocity correlation
functions are dependent on time differences and on differences of the coordinates $x$ and
$y$. Say, the eddy diffusion tensor (\ref{pe5}) is independent of time and does depend on
differences $x_1-x_2$ and $y_1-y_2$. However, it depends on both $z_1$ and $z_2$ due to
the strong inhomogeneity of the system in the direction perpendicular to the wall. A
$z$-dependence of the eddy diffusion tensor components can be found directly from the
proportionality laws $v_x,v_y\propto z$ and $v_z\propto z^2$. Say,
 \begin{eqnarray}
 D_{zz}(x,y,z_1;x,y,z_2)
 =\mu z_1^2 z_2^2 ,
 \label{mudef}
 \end{eqnarray}
where $\mu$ is a constant characterizing strength of the velocity fluctuations in the
peripheral region.

The equation for the first moment of $\theta$ (average value of the passive scalar
field), $\langle\theta\rangle$, is
 \begin{eqnarray}
 \partial_t\langle\theta\rangle=\partial_z
 \left[\mu z^4 \partial_z\langle\theta\rangle\right]
 +\kappa\partial_z^2\langle\theta\rangle ,
 \label{pe6}
 \end{eqnarray}
as it follows from Eqs. (\ref{pe10},\ref{mudef}). Comparing the advection and the
diffusion terms in the equation (\ref{pe6}) one finds a characteristic diffusion length
$r_{bl}$ defined as
 \begin{equation}
 r_{bl}=(\kappa/\mu)^{1/4}.
 \label{rbl}
 \end{equation}
The quantity determines the thickness of the diffusion boundary layer formed near the
wall. Due to smallness of $\kappa$ (remind that the Peclet or the Schmidt number is
assumed to be large) the diffusion length is much less than the thickness of the
peripheral region, where the law $v_z\propto z^2$ is satisfied.

We consider an advanced stage of the passive scalar decay or a statistically stationary
situation caused by the permanent passive scalar flux through the wall. Then the passive
scalar $\theta$ is non-zero primarily in the diffusive boundary layer, at $z\lesssim
r_{bl}$ (remind that after subtraction of its bulk value the field $\theta$ should tends
to zero in bulk that is at $z\to\infty$). However, we are interested mainly in the
passive scalar transport through the region $z\gg r_{bl}$, where the passive scalar is
carrying from the diffusive boundary layer to bulk. There it is possible to neglect the
molecular diffusion term in Eq. (\ref{pe6}) and we arrive at the proportionality law
 \begin{equation}
 \langle\theta\rangle \propto z^{-3} ,
 \label{first}
 \end{equation}
that gives the decaying rate of the average $\theta$ as $z$ grows. Note that the law
(\ref{first}) corresponds to a constant average passive scalar flow through the planes
$z=\mathrm{const}$, that is the flux is independent of $z$.

One anticipates that at $z\gg r_{bl}$ high passive scalar moments possess some scaling
behavior as well as $\langle\theta\rangle$. We introduce the corresponding scaling
exponents
 \begin{equation}
 \langle \theta^n \rangle
 \propto z^{-\eta_n},
 \label{expps}
 \end{equation}
If the molecular diffusion is irrelevant outside the diffusion boundary layer then
$\eta_n=3$ \cite{04LT}. However, our numerical data imply that the molecular diffusion is
relevant even at $z\gg r_{bl}$ (below we give an explanation of the phenomenon).
Therefore the exponents $\eta_n$ are not equal to $3$ and their values are subject of a
special investigation.

Let us turn to the passive scalar correlation functions (\ref{pe9}). At $z\gg r_{bl}$
their dependence on the coordinates along the wall are characterized by a correlation
length $l$, that can be found by balance of the molecular and the eddy diffusion along
the wall. The eddy diffusivity term in Eq. (\ref{pe10}) can be estimated as $\mu z^2$,
see Eq. (\ref{mudef}), the law $\propto z^2$ follows from the $z$-dependence of the
velocity components. Comparing the molecular diffusion term $\sim \kappa/l^2$ and the
eddy diffusion term in Eq. (\ref{pe10}), one finds
 \begin{equation}
 l\sim \sqrt{\kappa/\mu}\ z^{-1}.
 \label{corrl}
 \end{equation}
The quantity is of the order of $r_{bl}$ at $z\sim r_{bl}$ and diminishes $\propto
z^{-1}$ as $z$ grows.

\subsection{Integral Passive Scalar}

To exclude effects of the molecular diffusion, we introduce an integral of the passive
scalar field along a surface parallel to the wall
 \begin{equation}
 \Theta(t,z) = A^{-1}\int dx\ dy\
 \theta(t,x,y,z) ,
 \label{Thetadef}
 \end{equation}
where $A$ is the area of the surface and $z$ is its separation from the wall. We
designate correlation functions of the object as
 \begin{equation}
 \Phi_n(t,z_1,\dots,z_n)=
 \langle \Theta(t,z_1) \dots \Theta(t,z_n)\rangle \,.
 \label{corrf}
 \end{equation}
Integrating Eq. (\ref{pe10}) over $x_k$ and $y_k$, one obtains
 \begin{eqnarray}
 \partial_t\Phi_n
 =\kappa\sum_{m=1}^n\partial_{mz}^2\Phi_n
 +\int dx_1\dots dx_n dy_1 \dots dy_n
 \nonumber \\
 \left\{ \sum_{m,k=1}^n\! \! \partial_{mz}
 \left[D_{zz} \partial_{kz}F_n\right]
 +\sum_{m\neq k}\partial_{mz}\left[
 (\partial_{kz}D_{zz}) F_n \right] \right\} ,
 \label{pe100}
 \end{eqnarray}
where $D_{zz}=D_{zz}(\bm r_m,\bm r_k)$. At deriving Eq. (\ref{pe100}) we have taken some
integrals in part and used the constraint
 \begin{equation}
 \frac{\partial}{\partial x_k}D_{zx}(\bm r_m,\bm r_k)
 \!+\!\frac{\partial}{\partial y_k}D_{zy}(\bm r_m,\bm r_k)
 \!+\!\frac{\partial}{\partial z_k}D_{zz}(\bm r_m,\bm r_k)
 \!=\!0 ,
 \nonumber
 \end{equation}
which is a consequence of the incompressibility condition $\nabla\bm v=0$.

One expects a scaling behavior of the correlation functions (\ref{corrf}) at $z\gg
r_{bl}$. Then the last term in Eq. (\ref{pe100}) can be estimated as $\mu z^4
\partial_z^2\Phi$. Consequently, the term with the molecular diffusion $\kappa$ in Eq.
(\ref{pe100}) can be neglected in the region. The argumentation is the same as was used
for the moment $\langle\theta\rangle$, where the law (\ref{first}) was derived from Eq.
(\ref{pe6}). Note also, that in the decaying case the time derivative in Eq.
(\ref{pe100}) can be estimated as $\kappa/r_{bl}^2=\sqrt{\kappa\mu}$, the term is much
less than $\mu z^4\partial_z^2\sim \mu z^2$ at $z\gg r_{bl}$. Therefore the term with the
time derivative can be neglected in the region as well, and we obtain quasi-stationary
equations for $\Phi_n$. That reflects adiabaticity of the passive scalar statistics.

It is reasonably to assume that the correlation function $F_n(t,\bm r_1,\dots,\bm r_n)$
is correlated along the $X-Y$ plane on distances of the order of the correlation length
$l$ (\ref{corrl}) that is much smaller than the characteristic velocity length (width of
the peripheral region), then $D_{zz}$ in Eq. (\ref{pe100}) can be substituted by $\mu
z_m^2 z_k^2$. Then we obtain closed equations for the correlation functions
 \begin{eqnarray}
 \partial_t\Phi_n(t,z_1,\dots,z_n)=
 \mu\sum_{m,k=1}^n\frac{\partial}{\partial z_m}
 \left(z_m^2 z_k^2\frac{\partial}{\partial z_k}\Phi_n\right)
 \nonumber \\
 +2\mu \sum_{m\neq k}\frac{\partial}{\partial z_m}
 \left(z_m^2 z_k \Phi_n \right) , \qquad
 \label{correq}
 \end{eqnarray}
where we omitted the molecular diffusion term (see the above argumentation).

The equations (\ref{correq}) lead to the following closed equations for the moments of
the integral passive scalar
 \begin{equation}
 \partial_t \langle\Theta^n\rangle
 =\mu\left[z^4 \partial_z^2+4n z^3 \partial_z
 +4n(n-1)z^2\right]\langle\Theta^n\rangle \,.
 \label{moments}
 \end{equation}
In the stationary (or quasi-stationary) case (where $\partial_t\langle\Theta^n\rangle$ is
negligible) we obtain a homogeneous differential equation for the $n$-th moment that
admits a power solution
 \begin{equation}
 \langle\Theta^n\rangle\propto z^{-\zeta_n} \,.
 \label{scaling}
 \end{equation}
The exponents $\zeta_n$ can be easily found from Eq. (\ref{moments}), they are
 \begin{equation}
 \zeta_n=2n-1/2+\sqrt{2n+1/4}\,.
 \label{zeta}
 \end{equation}
We have chosen the positive sign of the square root leading to the reference value
$\zeta_1=3$, as it should be in accordance with Eq. (\ref{first}). We observe anomalous
scaling, that is a non-linear dependence of $\zeta_n$ on $n$, that can be compared with
the anomalous scaling of the passive scalar in the Kraichnan model
\cite{95GK,95SS,95CFKL}.

A natural conjecture that enables one to relate the moments of the passive scalar
$\theta$ and those of the integral passive scalar $\Theta$ is in using the passive scalar
correlation length $l$ as a recalculation factor. Then we come to an estimation
 \begin{equation}
 \langle \Theta^n \rangle\sim
 \frac{l^{(d-1)(n-1)}}{A^{n-1}}
 \langle \theta^n \rangle.
 \label{width}
 \end{equation}
Here $d$ is dimensionality of space, that is equal to $3$ in real flows but can be
arbitrary in numerical simulations. The estimation (\ref{width}) together with Eq.
(\ref{corrl}) lead to the relation
 \begin{equation}
 \eta_n=\zeta_n-(n-1)(d-1),
 \label{diff}
 \end{equation}
between the exponents introduced by Eqs. (\ref{expps},\ref{scaling}).

\section{Simulations}
\label{sec:simul}

We conducted Lagrangian simulations where dynamics of a large number of particles
subjected to flow advection and Langevin forces (producing diffusion) is examined. The
set of the particles is used instead of the passive scalar field $\theta$, that can be
interpreted as density of the particles. A big advantage of the approach is its
applicability to different space dimensions $d$. Indeed, in the scheme the number of
variables (coordinates of the particles) grows as a power of $d$ (at a given number of
the particles), but not exponentially.

To establish principal qualitative features of the passive scalar transport, we perform
mainly $2d$ simulations. The setup is periodic in $x$ (coordinate along the wall) with a
period $L$. In majority of simulations the velocity field is chosen to be
 \begin{eqnarray}
 v_x=z\left(\xi_1\cos\frac{2\pi x}{L}
 +\xi_2\sin\frac{2\pi x}{L}\right)\frac{L}{\pi} ,
 \label{vxxx} \\
 v_z=z^2\left(\xi_1\sin\frac{2\pi x}{L}
 -\xi_2\cos\frac{2\pi x}{L}\right) ,
 \label{vzzz}
 \end{eqnarray}
where $\xi_1$ and $\xi_2$ are independent random functions of time. Let us stress, that
the velocity field (\ref{vxxx},\ref{vzzz}) satisfies the incompressibility condition
$\partial_x v_x+\partial_z v_z=0$ for any functions $\xi_1(t)$ and $\xi_2(t)$. They are
chosen to possess identical Gaussian probability distributions, then the velocity field
(\ref{vxxx},\ref{vzzz}) has statistics homogeneous in $x$ (along the wall). In our
numerics, we have chosen $L=10$.

 \begin{figure}
 \centering
 \includegraphics[width=0.45\textwidth]{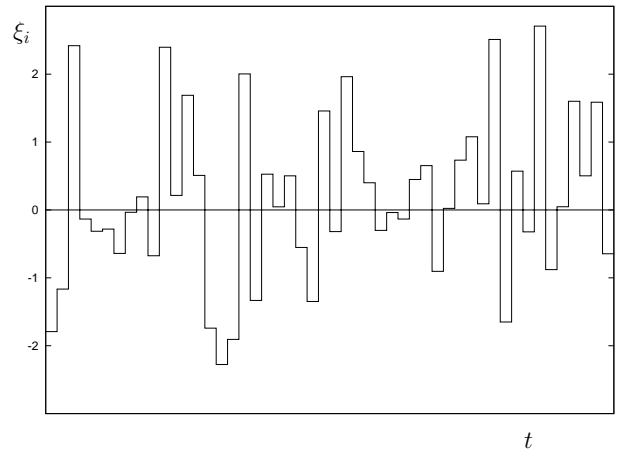}
 \caption{An example of the telegraph process.}
 \label{fig:telegr}
 \end{figure}

Though in reality the proportionality laws $v_x\propto z$ and $v_z\propto z^2$ are
satisfied inside the laminar boundary layer, in our setup the expressions for the
velocity components like (\ref{vxxx},\ref{vzzz}) are formally used at all $z$, that is
bulk corresponds to $z=\infty$. However, the law $v_z\propto z^2$ implies that the
particles may reach the $z$-infinity for a finite time. That ensures a finite particle
flux to bulk since sometimes a finite number of particles passages there. Therefore the
passive scalar transport to bulk is well defined in our setup.

Since the velocity correlation time in the peripheral region is much less than the
passive scalar mixing time, one should consider $\xi_1(t)$ and $\xi_2(t)$ as white
noises. However, in computer simulations one cannot realize zero correlation time. We
model the functions by telegraph processes, where both functions, $\xi_1$ and $\xi_2$,
remain constants inside time slots of a small duration $\tau$, and the values of $\xi_1$
and $\xi_2$ inside the slots are chosen to be random variables with identical normal
distributions. An example of such telegraph process is plotted in Fig.~\ref{fig:telegr}.
In our simulations the averages were $\langle\xi_1^2\rangle =\langle\xi_2^2\rangle=1$ and
different slot sizes were used, $\tau=0.001;0.002;0.004$. Then, in accordance with the
definition (\ref{mudef}), $\mu=\tau/2$.

In our scheme a particle trajectory $\bm\varrho(t)$ obeys the equation
 \begin{equation}
 \partial_t \bm\varrho=
 \bm v(t,\bm\varrho) + \bm\zeta(t),
 \label{particle}
 \end{equation}
where the first term represents the particle advection and the second term represents the
Langevin force. Let us stress that the variables $\bm\zeta$ are independent for different
particles whereas the variables $\xi_1$ and $\xi_2$ are identical for all particles,
according to physical meaning of the variables. The Langevin force $\bm\zeta$ is also
modeled by a telegraph process with the same time slot duration $\tau$ and with normal
distributions of the values in the slots. To ensure a given value of the diffusion
coefficient $\kappa$, one should choose
 \begin{equation}
 \langle \zeta_x^2 \rangle =\langle \zeta_z^2 \rangle=2\kappa/\tau.
 \label{langevin}
 \end{equation}
In majority of simulations we have chosen $\kappa=\tau/2$, and therefore the diffusive
length was $r_{bl}=1$, in accordance with the definition (\ref{rbl}).

Inside a time slot all the variables, $\xi_1$, $\xi_2$ and $\bm\zeta$, are
time-independent constants and the equation (\ref{particle}) becomes an autonomous
ordinary differential equation. It was solved as follows. A time slot was divided into a
number of time intervals and the equation was solved (without the Langevin force) using
the second order Runge-Kutta method. The number of intervals is $z$-dependent, it is
proportional to $z$ at $z>2.5$. For $z>12$ we solved equations for $1/\varrho_z$ instead
of $\varrho_z$. Both features are motivated by the strong dependence of the velocity on
$z$, $v_z\propto z^2$. After solving the equation inside a slot a term produced by the
Langevin force was added. To examine role of diffusion outside the diffusive boundary
layer, in some simulations we switched off the diffusion (the Langevin forces) at
distances $z>z_d$ (where a choice of $z_d$ is different for different cases).

The particles are permanently injected near the wall in random positions at the beginning
of each time slot. The simulations were performed in the stripe $0<z<100$, the particles
crossing the lines $z=0$ and $z=100$ were excluded from the set. The number of particles
leaving the region $0<z<100$ through the wall is much larger than the number of particles
escaping through the line $z=100$. The last ones correspond to the passive scalar
transport to bulk. A balance between the particle injection and losses leads to a
statistical equilibrium achieved gradually in the numerics. Thus, our simulations cover
the statistically stationary passive scalar transport. It corresponds both to the steady
temperature distribution supported by a constant heat flux from the wall and to the decay
of the concentration of pollutants that can be treated adiabatically.

An extension of our scheme to higher dimensions, $d>2$, is straightforward. We utilize
the same equation (\ref{particle}) where all quantities have $d$ components. The Langevin
forces $\bm\zeta$ are determined by the same relations (\ref{langevin}) and a
generalization of the expressions (\ref{vxxx},\ref{vzzz}) is as follows. The velocity
$\bm v$ is determined by a set of $2(d-1)^2$ random variables $\xi_{1ij}$ and
$\xi_{2ij}$:
 \begin{eqnarray}
 v_i=z\sum\limits_{j=1}^{n-1}\left(\xi_{1ij}\cos\frac{2\pi x_j}{L}
       +\xi_{2ij}\sin\frac{2\pi x_j}{L}\right) \frac{L}{\pi},
 \label{vxxj} \\
 v_z=z^2\sum\limits_{j=1}^{n-1}\left(\xi_{1jj}\sin\frac{2\pi x_j}{L}
       -\xi_{2jj}\cos\frac{2\pi x_{j}}{L}\right),
 \label{vzzj}
 \end{eqnarray}
where the subscripts $i$, $j$ numerate first $d-1$ space coordinates and the last $d$-th
coordinate is $z$. Here, all $\bm\zeta$, $\xi_{1ij}$ and $\xi_{2ij}$ are, again,
telegraph processes with the same statistical properties as above, and we use the
second-order Runge-Kutta scheme inside a time slot.

In terms of the particles, the passive scalar field $\theta$ is defined as a number of
particles per unit volume, that is as a number of particles inside a box divided by the
box volume. Of course, the definition works well provided the box is small (in comparison
with all characteristic scales of the problem) and the number of particles inside the box
is large. To satisfy these contradictory conditions one should deal with a large enough
total number of particles. That is why the injection rate in our numerics is chosen to
produce a large number of particles, $10^5\div10^6$, in the statistical equilibrium.

\subsection{Tongues}

Our simulations shown that the passive scalar transport to bulk is related to specific
structures of the passive scalar. The passive scalar is concentrated mainly in the narrow
diffusive layer near the wall. However, sometimes a fluid jet is generated carrying the
passive scalar from the wall towards bulk, that produces a passive scalar tongue with
width (cross-section) diminishing as $z$ grows. The property is a consequence of the law
$v_z\propto z^2$ reading that the $z$-component of the tongue velocity increases as $z$
grows. Thus we come to a geometrical interpretation of the passive scalar correlation
length $l$: it is the characteristic size of the tongue cross-section (taken along the
wall). The cross-section behavior corresponds to an expected decrease of the correlation
length as $z$ grows. Let us stress that in accordance with Eq. (\ref{corrl}) the
characteristic tongue cross section is dependent on the diffusion coefficient $\kappa$.

\begin{figure}
 \centering
 \includegraphics[width=0.45\textwidth]{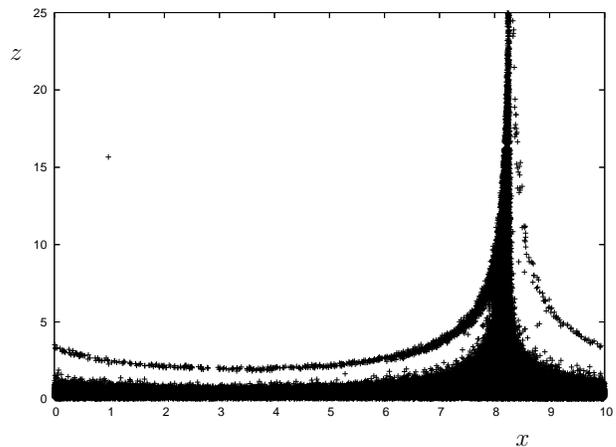}
 \caption{An example of a passive scalar structure formed near the wall in
 the $2d$ random flow. Different particles are designated by small crosses.}
 \label{fig:tongue2}
 \end{figure}

A tongue is typically pulled from a ``bump'' of the passive scalar distribution. After
some time the tongue is tilted and then pressed back to the diffusive layer. Then next
tongue is pulled, usually from the bump remaining at bottom of the previous tongue, and
is, in turn, pressed back to the diffusive layer. As a result, a complicated multi-fold
structure is formed, an example of such structure is drawn in Fig. \ref{fig:tongue2},
that represents a snapshot generated in our simulations.

\begin{figure}
 \centering
 \includegraphics[width=0.45\textwidth]{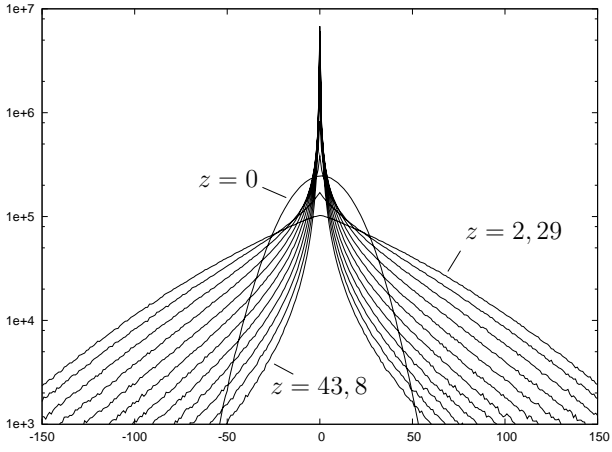}
 \caption{Histograms of the passive scalar flux at different separations from the wall.
 The root mean square fluctuations are much larger than the average value and the
 histograms are practically symmetric. At $z=0$ the probability distribution is Gaussian
 whereas at $z>r_{bl}$ it has exponential tails.}
 \label{fig:gist1}
 \end{figure}

Sometimes the tongue is pulled upto $z$-infinity, and then a portion of the passive
scalar (a number of particles) is pushed to bulk. After that the tongue is tilted and the
passive scalar current to bulk stops. That implies that the passive scalar flux, $\int
dx\ \theta v_z$ in $2d$, is highly intermittent quantity at $z>r_{bl}$. The conclusion is
confirmed by the flux histograms drawn in Fig.~\ref{fig:gist1} for different $z$. In
simulations, the passive scalar flux was measured as a number of particles crossing the
plain $z=\mathrm{const}$ during a time $\tau$. At $z=0$ the flux probability distribution
is practically Gaussian, being formed by a balance between the random injection of the
particles and their leaving the wall. However, the distribution becomes less and less
Gaussian as $z$ grows. The histograms in Fig.~(\ref{fig:gist1}) are practically
symmetric. The property is related to the fact that only a small amount of particles in
the tongue are pulled to bulk, the majority of particles returns back, that produces
practically equal fluxes to bulk and toward the wall. That explains why at $z\gg r_{bl}$
the root mean-square fluctuation of the flux is much larger than its average value.

\subsection{Moments}

Based on numerical data one can compute moments and correlation functions of different
quantities characterizing the passive scalar statistics. One can consider both local
functions or integral objects. All the quantities are computed as time averages.

 \begin{figure}
 \centering
 \includegraphics[width=0.45\textwidth]{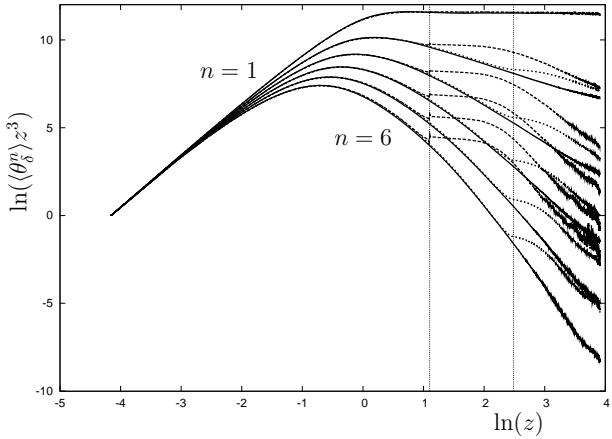}
 \caption{Log-log plot of the moments of $\theta_\delta$,
 $\langle\theta_\delta^n\rangle$, multiplied by $z^3$, for $\delta=0.03125$
 and $n=1\div 6$. The graph reflects simulations where diffusion occurs
 everywhere, and is switched off at $z=3$ or $z=12$.}
 \label{fig:zkub}
 \end{figure}

We introduce an object $\theta_\delta$ that is a number of particles inside a square box
of size $\delta$ divided by its area $\delta^2$ (in $2d$). The quantity $\theta_\delta$
is close to $\theta$ provided the number of particles is large and the size of the box is
small enough. Moments $M_n$ of the quantity $\theta_\delta$,
$M_n=\langle\theta_\delta^n\rangle$, for $n=1\div 6$ are computed as averages over time
intervals $10^6\div 10^7 \tau$ as $\delta=0.03125$. The results are presented in
Fig.~\ref{fig:zkub} where the moments multiplied by $z^3$ are plotted in log-log
coordinates (solid curves). We see that the prediction (\ref{first}) for the first moment
is perfectly satisfied whereas higher moments deviate strongly from the diffusionless law
$\propto z^{-3}$. We conclude that the diffusion is relevant at $z>r_{bl}$, indeed.

\begin{figure}
 \centering
 \includegraphics[width=0.45\textwidth]{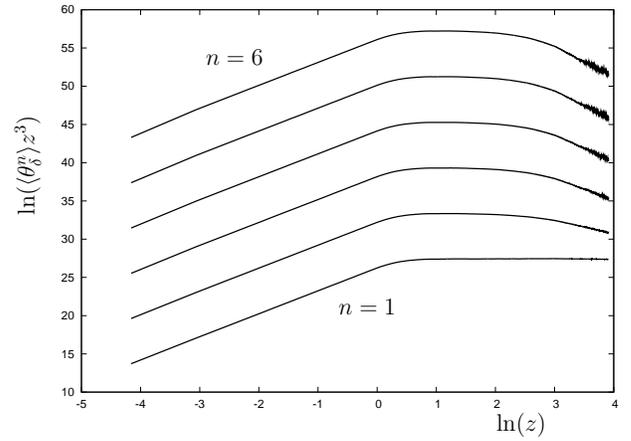}
 \caption{Log-log plot of the moments of $\theta_\delta$,
 $\langle\theta_\delta^n\rangle$, multiplied by $z^3$, for $\delta=0.03125$
 and $n=1\div 6$ in the case where diffusion is substituted by a constant velocity
 carrying the particles from the wall.}
 \label{fig:ckub}
 \end{figure}

To verify the conclusion we repeated the simulations switching off diffusion at $z>3$ and
at $z>12$. Results are presented in Fig.~\ref{fig:zkub} by dashed curves. We see an
appearance of plateaus, starting just from $z=3$ or $z=12$ and corresponding to the law
$\propto z^{-3}$, in accordance with Ref. \cite{04LT}. The plateaus are observed in
restricted regions of separations from the wall $z$ slightly diminishing as $n$ grows. An
explanation of the fact is that cutoffs of the plateaus are observed where $\delta$
becomes of the order of the passive scalar correlation length (along the wall). To check
this conjecture we repeated the simulations for larger values of $\delta$ and observed
that the plateaus shrink as $\delta$ grows. That confirms our explanation. To be
absolutely sure that the diffusion is relevant we performed simulations without diffusion
but with adding a constant velocity $V$ carrying the particles away the wall. Results of
the simulations are presented in Fig.~\ref{fig:ckub} where the moments
$\langle\theta_\delta^n\rangle$, multiplied by $z^3$, are plotted. We see plateaus
signalling that outside the boundary layer the passive scalar moments behave in
accordance with the diffusionless prediction.

\begin{figure}
 \centering
 \includegraphics[width=0.45\textwidth]{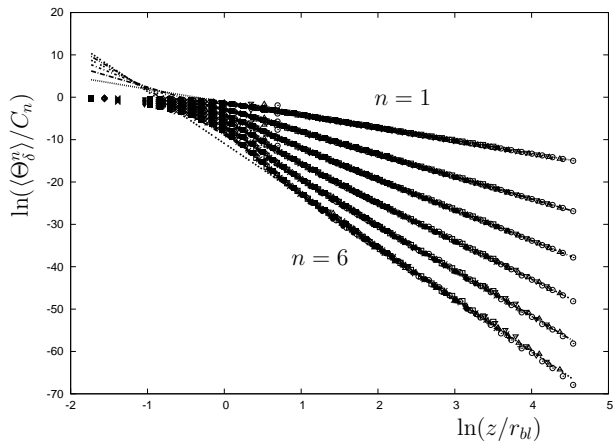}
 \caption{Moments of $\Theta_\delta$ in log-log coordinates, $n=1\div6$.
 In the region $z>r_{bl}$ the results collapse onto single curves for
 three times $\tau = 0.001, 0.002, 0.004$ and four different values of the
 diffusion coefficient.}
 \label{fig:lines}
 \end{figure}

The next object of our investigation is the integral quantity $\Theta$ that is the
passive scalar integrated along the wall, see the definition (\ref{Thetadef}). In
numerics it is determined by a number of particles in a slice of thickness $\delta$,
parallel to the wall, divided by its volume (area), we designate the ratio as
$\Theta_\delta$. In our $2d$ setup the area is equal to $L\delta$, where $\delta$ is
chosen to be much less than $z$. The moments of $\Theta_\delta$, $\langle\Theta_\delta^n
\rangle$, are computed by time averaging over a long time $\sim 10^7\tau$. To check
robustness of the results we performed computations for different time slots, $\tau =
0.001, 0.002, 0.004$, and for four different values of the diffusion coefficient
$\kappa$. The figure \ref{fig:lines} demonstrates that the values of each moment collapse
to a single curve in the logarithmic coordinates $\ln(z/r_{bl})$ and $\ln(\langle
\Theta^n \rangle/C_n)$ where the factors $C_n$ are the corresponding moments near the
wall.

\begin{figure}
 \centering
 \includegraphics[width=0.45\textwidth]{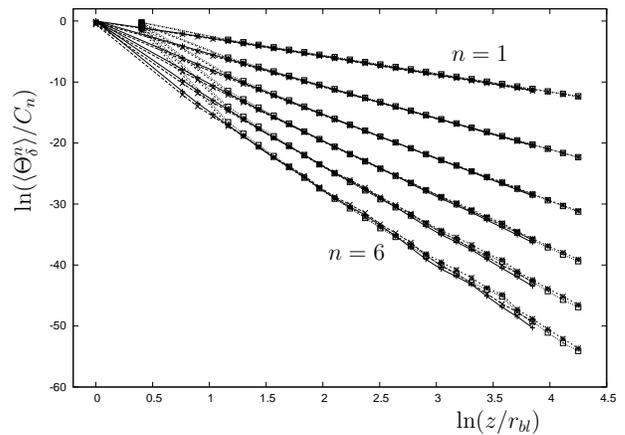}
 \caption{Moments of $\Theta_\delta$ in log-log coordinates, $n=1\div6$. The results
 are obtained for two cases where the diffusion occurs everywhere and where it is
 switched off at $z>3$, and also for two different velocity fields: with two and four
 harmonics.}
 \label{fig:diffuz}
 \end{figure}

One can check that, in accordance with our theoretical expectations, the moments of
$\Theta_\delta$ are insensitive to diffusion. To illustrate the assertion we present in
Fig.~\ref{fig:diffuz} the moments of $\Theta_\delta$ computed at $\tau=0.002$ for two
cases: in the first case the diffusion occurs everywhere and in the second case it is
switched off at $z>3$. One can observe no difference between the data. One can also try
to use an alternative to the expression (\ref{vxxx},\ref{vzzz}) velocity field. There are
two series of data plotted in Fig.~\ref{fig:diffuz} and corresponding to the velocity
(\ref{vxxx},\ref{vzzz}) and to the velocity with four random factors
  \begin{eqnarray}
  \frac{\pi v_x}{L z}
  =\xi_1\cos\frac{2\pi x}{L}
  +\xi_2\sin\frac{2\pi x}{L}
  \nonumber \\
  +\xi_3\cos\frac{4\pi x}{L}
  +\xi_4\sin\frac{4\pi x}{L},
    \label{vxxc}       \\
  \frac{v_z}{z^2}
  =\xi_1\sin\frac{2\pi x}{L}
  -\xi_2\cos\frac{2\pi x}{L}
  \nonumber \\
  +2\xi_3\sin\frac{4\pi x}{L}
  -2\xi_4\cos\frac{4\pi x}{L} .
 \label{vzzc}
 \end{eqnarray}
Again, there is no visible difference between the series.

\begin{figure}
 \centering
 \includegraphics[width=0.45\textwidth]{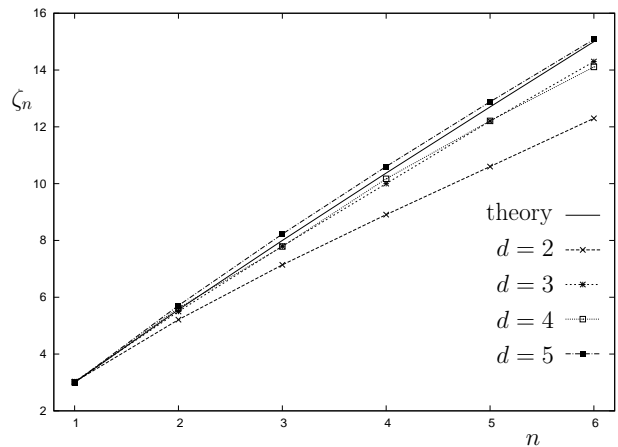}
 \caption{Exponents of the moments $\langle\Theta_\delta^n\rangle$,
 for $n=1\div 6$ and space dimensions $d=2\div 5$. For comparison the theoretical
 curve $\zeta_n=2n-1/2+\sqrt{2n+1/4}$ is plotted (solid line).}
 \label{fig:difference}
 \end{figure}

We observe that the moments of $\Theta$ are decreasing functions of $z$ that are
power-like in the region $z>r_{bl}$. Extracting the scaling exponents $\zeta_n$, see the
definition (\ref{zeta}), for $n=1\div 6$ in $2d$ we obtain values that are presented in
Fig. \ref{fig:difference} as the lower set of points (some smooth curve is drawn through
the points for better visualization). We conducted analogous simulations for higher
dimensions, upto $d=5$. The results are depicted in the same figure \ref{fig:difference}.
We see that the exponents $\zeta_n$ depend on $d$, however, for $d\geq3$ they are close
to the theoretical values (\ref{zeta}) represented by a solid line.

One can think that the deviations from the theoretical values (\ref{zeta}) are related to
an existence of additional passive scalar (relatively long) correlations along the wall
that can be produced by the multi-fold structures of the type drawn in
Fig.~\ref{fig:tongue2}. The long correlations should lead to increasing moments of the
passive scalar in comparison with the short correlated case. It is naturally to expect
that the fold effect becomes less pronounced in higher dimensions. Indeed,
Fig.~\ref{fig:difference} shows that the deviations from the values (\ref{zeta}) diminish
as the space dimensionality $d$ grows. That confirms our explanation.

\begin{figure}
 \centering
 \includegraphics[width=0.45\textwidth]{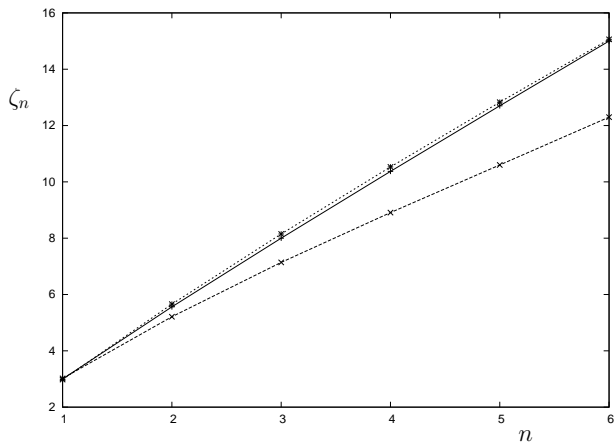}
 \caption{Exponents of the moments $\langle\Theta_\delta^n\rangle$,
 for $n=1\div 6$ and space dimensions $d=2$ for two different velocity
 fields: containing only first harmonic and three (ninth, tenth and eleventh)
 harmonics. For comparison the theoretical curve $\zeta_n=2n-1/2+\sqrt{2n+1/4}$
 is plotted (solid line).}
 \label{fig:complexflow}
 \end{figure}

To check our conjecture, we conduct simulations for the velocity field containing, like
the expressions (\ref{vxxc},\ref{vzzc}), a set of harmonics in terms of the period $L$:
the ninth, tenth and eleventh ones. In such velocity field correlations related to the
multi-fold tongue structures have to be suppressed, and, consequently, the exponents
$\zeta_n$ should be close to the theoretical values (\ref{zeta}). This expectation is
confirmed by our simulations, the results are presented in Fig. \ref{fig:complexflow},
where measured exponents are plotted. We see a good agreement of the measured and
theoretical exponents.

\begin{figure}
 \centering
 \includegraphics[width=0.45\textwidth]{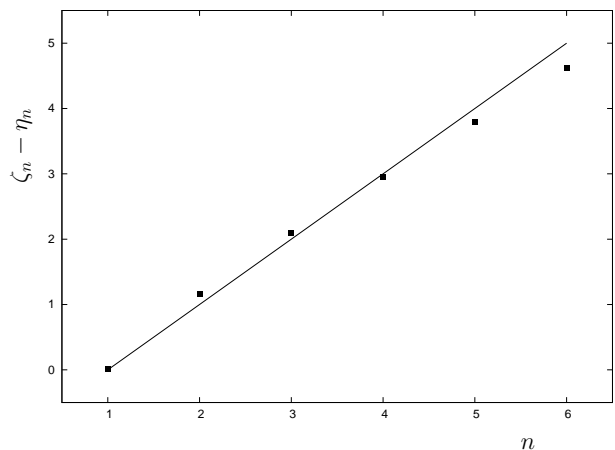}
 \caption{The difference of the scaling exponents of moments for the integral passive
 scalar and for the passive scalar, $\zeta_n-\eta_n$, computed at $\delta=0.03125$ in $2d$.
 For comparison, the theoretical prediction $n-1$ is drawn.}
 \label{fig:corrlength}
 \end{figure}

One can extract from our numerical data the exponents $\eta_n$ of the moments of
$\theta_\delta$, see the definition (\ref{expps}), as well as $\zeta_n$. It is
interesting to check the theoretical prediction (\ref{diff}). For the purpose we plotted
the difference $\zeta_n-\eta_n$ as a function of $n$, see Fig. \ref{fig:corrlength}. For
comparison, the theoretical straight line $n-1$ (correct in $2d$) is drawn in the same
figure. We see a good agreement, confirming the scaling (\ref{corrl}) of the passive
scalar correlation length $l(z)$ along the wall.

\section{Conclusion}
\label{sec:conc}

We performed extensive numerical simulations of the passive scalar mixing in peripheral
regions of random flows like high-Reynolds turbulence. The simulations confirm earlier
theoretical expectations and reveal a lot of new details. At advanced stages of the
passive scalar mixing passive scalar fluctuations are concentrated mainly in the narrow
diffusive layer near the boundary provided the Peclet or the Schmidt number is large. We
found that the passive scalar transport from the diffusive boundary layer to bulk is
related to passive scalar tongues formed by jets directed to bulk. The tongues are
objects responsible for strong intermittency characteristic of the passive scalar
transport through the peripheral region.

We examined the passive scalar statistics outside the diffusive boundary layer and
realized that both, moments of the passive scalar $\theta$ and of the passive scalar
integrated along the wall, $\Theta$, possess well pronounced scaling in terms of the
separation from the wall $z$. We compared the corresponding exponents extracted from our
numerics with our theoretical scheme and established their agreement. However, one should
be careful since our theoretical predictions are correct for an infinite vessel and can
be violated for the numerics where the velocity correlation length along the wall
coincide with the velocity period. We found also an agreement between the theoretical
prediction for the tongue cross-section dependence on $z$ and our numerics. Therefore the
simulations confirm our theoretical predictions.

There remain some problems to be solved in future. We are going to extend our
consideration incorporating average flows (like in pipes) which are shear-like near the
wall. Another natural extension of our approach is related to chemical reactions in
random flows. One can also note polymer solutions, where the polymer elongation is very
sensitive to the character of the flow. The problem is significant, e.g., for the elastic
turbulence. However, a long-time memory characteristic of the polymer solutions could
modify our results. We considered smooth walls in our work. There is a set of questions
related to roughness of the wall, possible corners, caverns and peaks. All the objects
can modify our conclusions, it is a subject of special investigation.

Our results agree qualitatively with data known from investigations of turbulent plumes
in turbulent flows where a complicated space structure of the passive scalar fluctuations
is observed \cite{00VSAW,02CWK,06CK,06CKM,07DSW}. We believe that statistical properties
of the structure can be explained on the basis of our results implying production of the
passive scalar tongues pushing to bulk. The explanation needs a generalization of our
scheme where turbulent velocity fluctuations in bulk should be included. That is a
subject of future work.

\acknowledgements

We thank M. Chertkov, I. Kolokolov, V. Steinberg and K. Turitsyn for numerous helpful
discussions. Simulations were performed on the cluster Parma at the Landau Institute for
theoretical physics RAS and the NGU cluster. The work was partly supported by RFBR grant
09-02-01346-a and by Russian Federal target program Kadry.

\end{document}